\documentclass[%
reprint,
 amsmath,amssymb,
 aps,
]{revtex4-2}

\usepackage{float}
\usepackage{slashed}
\usepackage{xcolor}
\usepackage{amsmath}
\usepackage{subcaption}
\usepackage[english]{babel}
\usepackage{lineno, blindtext}
\usepackage{tikz}
\usepackage{adjustbox}
\usepackage{rotating}
\usepackage{rotfloat}
\usepackage{verbatim}
\usepackage[utf8]{inputenc}
\usepackage{graphicx}
\usepackage{dcolumn}
\usepackage{bm}
\usepackage[super]{nth}
\usepackage{lipsum}

\newcommand\Tdiag[4]{%
    \multicolumn{1}{|p{#2}|}{\hskip-\tabcolsep
    \begin{tikzpicture}[%
                baseline={(0,-.25\baselineskip)},
                every node/.style={outer sep=0pt,inner sep=#1}]
    \node[minimum width={#2+1\tabcolsep-\pgflinewidth},
        minimum height=2\baselineskip-\pgflinewidth+\extrarowheight,
        use as bounding box] (box) {};
    \draw[line cap=round] (box.north west) -- (box.south east);
    \node[anchor=south west,text width=.75*#2,align=left] at (box.south west) {#3};
    \node[anchor=north east,text width=.75*#2,align=right] at (box.north east) {#4};
\end{tikzpicture}\hskip-\tabcolsep}}

\usepackage{hyperref}
\usepackage{xcolor}
\hypersetup{
  colorlinks   = true, 
  urlcolor     = red, 
  linkcolor    = blue, 
  citecolor   = blue 
}

\usepackage{threeparttable}
\usepackage{mathtools}
\usepackage{siunitx}
\DeclarePairedDelimiterXPP\BigOSI[2]%
  {\mathcal{O}}{(}{)}{}%
  {\SI{#1}{#2}}

\begin{document}

\preprint{APS/123-QED}

\title{Search for the production of dark gauge bosons in the framework of Einstein-Cartan portal in the simulation of proton-proton collisions at $\sqrt{s} = 13.6$ TeV}

\author{S. Elgammal}
 \altaffiliation[sherif.elgammal@bue.edu.eg]{}
\affiliation{%
Centre for theoretical physics, The British University in Egypt. 
}%


\date{\today}

\begin{abstract}
{In this study, we investigate the potential production of a heavy torsion field (TS) at the LHC, which stems from a simplified model rooted in Einstein-Cartan gravity, in connection with dark matter. Within this framework, the torsion field is capable of decaying into pairs of dark matter (DM) particles. Notably, one of these DM particles is heavy enough to decay into dark neutral gauge bosons (A$^{\prime}$) alongside another DM particle.
The Analysis has been performed by studying events with dimuon plus missing transverse energy produced in the simulated proton-proton collisions at the Large Hadron Collider, at 13.6 TeV center of mass energy and integrated luminosity of 52 fb$^{-1}$ corresponding to the LHC run 3 circumstances during 2022 and 2023.
We provide upper limits, in case no new physics has been discovered, on the masses of various particles in the model as (A$^{\prime}$), as well as the heavy mediator (torsion field).}

\vspace{0.75cm}
\end{abstract}

\maketitle





\section{Introduction}
\label{sec:intro}
The Standard Model of particle physics SM has been tested for more than 40 years \cite{SMcource}, and its predictions agree very well with all experimental observations. 
However, the SM is nowadays considered as a low-energy manifestation of other theories realized at high energy, generically known as BSM (Beyond the Standard Model) theories \cite{SMandBSM}. 
One motivation for BSM physics is to have a unified theory for the electromagnetic, weak, and strong interactions, in a unique Grand Unified Theory (GUT) \cite{Extra-Gauge-bosons}. 
The Super-Symmetry (SUSY) attempts to also include gravitation, leading to models with extra spatial dimensions. These BSM models typically predict the
existence of new dark particles at the TeV scale and higher.

The existence of heavy neutral bosons (Z$^{\prime}$) is a feature of many extensions of the Standard Model. 
They arise in extended gauge theories, including grand unified theories (GUT) \cite{gaugeboson1}, and other models like left-right symmetric
models (LRM) \cite{LR-symmetry11}. A specific case is the sequential standard model (SSM), in which the Z$^{\prime}$ boson has the same coupling as the SM Z boson \cite{Super-symmetry12}. 
Model of extra dimensions like the Randall and Sundrum model (RS) \cite{extra-dimension2} predicts the existence of heavy Kaluza-Klein gravitons.
Searches for these heavy dark neutral gauge bosons have been performed at the CMS and ATLAS experiments, at the Large Hadron Collider (LHC), with no evidence of their existence using the full RUN II period of the LHC data taking \cite{zprime, zprimeATLAS}.

Another alternative for the Randall and Sundrum model could be achieved through the Einstein-Cartan (EC) portal \cite{EC1, EC2, EC3, EC4, EC5, EC6, EC7}. 
At which gravity (represented by torsion field) can couple to the SM particles in addition to dark sector fermions, it provides a mechanism of producing the dark sector particles and allows a chance for probing dark gauge boson (A$^{\prime}$), which corresponds to a $U(1)_{D}$ symmetry, at LHC \cite{R1}. 
In this theory, the torsion mass is in the TeV-scale regime, so that the A$^{\prime}$ can be produced with a high boost and missing transverse energy ($E^{miss}_{T}$) from dark-sector fermions. The search for the A$^{\prime}$ could be achieved at the LHC via its decay to dilepton (i.e. A$^{\prime} \rightarrow l^{+} l^{-}$) and large $E^{miss}_{T}$.

Many searches for DM have been performed by analyzing the data collected by the CMS experiment during RUN II. These searches rely on the production of a visible object "X", which recoils against the large missing transverse momentum from the dark matter particles, leaving a signature of ($\text{X} + E^{\text{miss}}_{T}$) in the detector \cite{R38}. 
The visible particle could be an SM particle like Z \cite{monozCMS,monozATLAS}, W
bosons \cite{monowATLAS1,monowATLAS2} or jets \cite{monojetATLAS,monojetCMS}, 
photon \cite{monophotonCMS,monophotonATLAS} or SM Higgs boson \cite{monohiggsCMS,monohiggsATLAS}. 

The ATLAS and CMS collaborations have previously conducted searches for the massive extra neutral gauge boson Z$^{\prime}$ decaying to dileptons. These searches are based on the Grand Unified Theory (GUT) and Supersymmetry \cite{Extra-Gauge-bosons, gaugeboson1, LR-symmetry11, Super-symmetry12}. 
Despite extensive efforts using the LHC data from the full RUN II period, no evidence of the existence of Z$^{\prime}$ has been found \cite{zprime, zprimeATLAS}.
Results from the CMS collaboration have excluded the existence of Z$^{\prime}$ in the mass range between 0.6 and 5.15 TeV with 95\% Confidence Level (CL), while the ATLAS experiment has excluded the mass range between 0.6 and 5.1 TeV.

The direct detection of dark matter in underground experiments \cite{XENON10, XENON100, XENON1T, SENSEI, PandaX-II} presents various constraints. The model suggests that the interaction between dark matter and electrons can effectively explore light-dark matter. Exclusion limits confidently eliminate the possibility of dark matter with masses exceeding 4 MeV \cite{DMdirect}.

Indirect detection searches for dark matter are currently very exciting. 
The annihilation of dark matter into leptons or hadrons in space can potentially produce a signal of thermal WIMPs. The most reliable constraints for GeV dark matter annihilation into various final states are provided by Fermi-LAT observations of dwarf spheroidal galaxies \cite{fermi1,fermi2}, AMS-02 positron fluxes \cite{AMS1,AMS2}, and CMB \cite{Planck111, CMB1, CMB2} limits. These results indicate that the WIMP dark matter mass window from 20 GeV up to the unitarity limit \cite{Unitarity} of 100 TeV remains largely open \cite{indirect1,indirect2}.

In this analysis, we present a search for dark neutral gauge bosons (A$^{\prime}$), which originated in a simplified model in the Einstein-Cartan portal, at the LHC simulated proton-proton collisions with 13.6 TeV center of mass energy corresponding to the LHC run 3 circumstances \cite{LumiPublicResults}. 
The topology of the studied simulated events is dimuon, from the decay of A$^{\prime}$, plus large missing transverse energy attributed to dark matter. 
The signal of the above process is very similar to that coming from the decay of 
Charginos pair into the lepton pair plus MET, carried out by ATLAS \cite{charginosATLAS}. 

In the following section \ref{section:model}, the theoretical formalism of the 
$U(1)_{D}$ simplified model based on the Einstein-Cartan gravity and its free parameters are presented. Then the simulation techniques used for event generation for the signal and SM background samples are displayed in section \ref{section:MCandDat}. 
Afterward, the selection cuts and the strategy of the analysis are explained in section \ref{section:AnSelection}. 
Finally, this analysis's results and summary are discussed in sections \ref{section:Results} and \ref{section:Summary}, respectively.

\begin{figure} [h!]
\centering
\resizebox*{6.0cm}{!}{\includegraphics{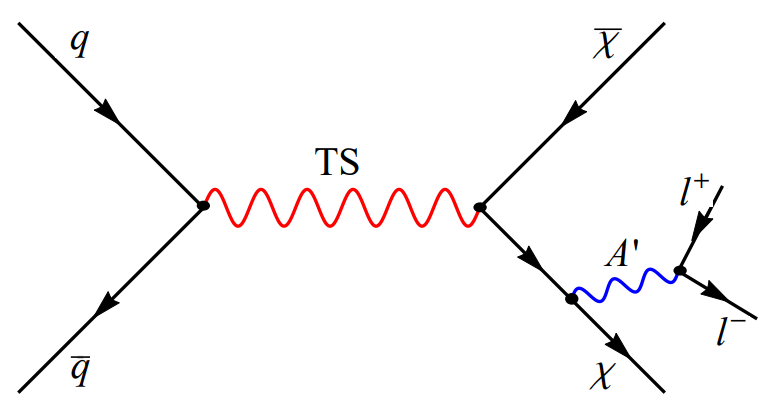}}
\caption{Feynman diagram for the simplified model based on Einstein-Cartan gravity; for the production of dark gauge boson (A$^{\prime}$) in association 
to dark matter ($\chi$) pair \cite{R1}.}
\label{figure:fig1}
\end{figure}

\section{The simplified model in the framework of Einstein-Cartan gravity}
\label{section:model}
The analyzed simplified model is based on the Einstein-Cartan gravity, which has been discussed in \cite{R1}, assumes the production of dark matter from proton-proton collisions at the LHC in addition to a new heavy neutral dark gauge boson A$^{\prime}$. 

The proposed dark gauge boson (A$^{\prime}_{\mu}$) can be produced through the process of pair annihilation of two quarks $q\bar{q}$ mediated by the heavy torsion field ($S_{\rho}$, which is the axial–vector part of the torsion tensor
$T^{\lambda}_{\mu\nu}$ \cite{R1}), which then undergoes two dark matter particles ($\chi$). 
Dark matter is heavy enough to decay to an A$^{\prime}_{\mu}$ and another dark matter ($\chi$) as shown in figure \ref{figure:fig1}.

The interaction terms, in the effective Lagrangian, between the torsion field 
and Dirac fermion $(\psi)$, are given by \cite{R1} 
\begin{equation}
\bar{\psi}i\gamma^{\mu}(\partial_{\mu} + i\texttt{g}_{\eta}\gamma^{5}S_{\mu} + ...)\psi. \nonumber
\end{equation}
According to the previous equation, the torsion field interacts with both the Standard Model (SM) and hidden fermions through its axial-vector part, with a universal coupling of $\texttt{g}_{\eta}$ = 1/8 = 0.125 \cite{R1} at the classical level. This indicates that despite the SM and dark sectors existing in separate realms, the torsion field's universal coupling to any fermion is essential for generating dark-sector fermions. This occurs when SM fermions collide, leading to the subsequent radiation or decay of lighter dark-sector particles, producing dark gauge bosons with a significantly unsuppressed production cross-section.

While the term in the effective Lagrangian at which the torsion field couples 
to the dark matter, and between dark gauge boson (A$^{\prime}_{\mu}$) and dark matter are given by \cite{R1} 
\begin{equation}
\bar{\chi}(i\gamma^{\mu}D_{\mu} - M_{\chi})\chi, \nonumber
\end{equation}
where $D_{\mu} = \partial_{\mu} + i\texttt{g}_{\eta}\gamma^{5}S_{\mu} + i\texttt{g}_{D} A^{\prime}_{\mu}$, $M_{\chi}$ is the dark matter mass and $\texttt{g}_{D}$ is the coupling of dark gauge boson to dark matter.
\begin{table*} 
\centering
\fontsize{10.pt}{12pt}
\selectfont
\begin {tabular} {lll}
\hline
\hline
Parameters & \hspace{1pt} Description & \hspace{1pt} optimized value \\
\hline
$\texttt{g}_{\eta}$  & Coupling of torsion field to Dirac fermions & \hspace{10pt} \\
$\texttt{g}_{D}$ & Coupling of dark gauge boson to dark matter & \hspace{10pt} \\
$M_{\chi}$ & Mass of dark matter $(\chi)$ & \hspace{10pt} $500$ GeV \cite{indirect2, Nature-indirect}\\
$M_{A'}$  & Mass of dark gauge boson $(A')$ & \hspace{10pt} $[200,...,900]$ GeV\\
$M_{TS}$  & Mass of torsion field $(TS)$ & \hspace{10pt} $[1250,...,7000]$ GeV\\
\hline
Mass assumption & \hspace{4pt} $M_{A^{\prime}} < 2M_{\chi}$ \cite{R1} \\
\hline
\hline
\end {tabular}
\vspace{3pt}
\caption{The simplified model parameters based on Einstein Cartan gravity and the chosen mass assumption \cite{R1}.}
\label{table:par}
\end{table*}
\begin{sidewaystable} []
\centering
\tiny
\scriptsize
\fontsize{8.pt}{12pt}
\selectfont
\begin{tabular}{|c|c|c|c|c|c|c|c|c|c|c|c|}
\hline
\Tdiag{.2em}{1.1cm}{$M_{A'}$}{$M_{TS}$}  & 1250 &1500 &1750 & 1800 & 1970 & 2000 & 3000 & 4000 & 5000 & 6000 & 7000 \\
\hline
200 & 4.3$\times10^{-4}$ &
    86.5$\times10^{-4}$ &
    158.0$\times10^{-4}$ &
    170.0$\times10^{-4}$ &
    180.0$\times10^{-4}$ &
    181.0$\times10^{-4}$ &
    89.5$\times10^{-4}$ &
    26.0$\times10^{-4}$ &
    6.1$\times10^{-4}$ &
    1.2$\times10^{-4}$ &
    2.3$\times10^{-5}$ \\
\hline
300  & 9.8$\times10^{-7}$ & 
    40.0$\times10^{-4}$ &
    110.0$\times10^{-4}$ &
    120.0$\times10^{-4}$ &
    142.0$\times10^{-4}$ &
    145.0$\times10^{-4}$ &
    87.4$\times10^{-4}$ &
    27.0$\times10^{-4}$ &
    6.6$\times10^{-4}$ &
    1.4$\times10^{-4}$ &
    2.6$\times10^{-5}$ \\
\hline
400  & 5.4$\times10^{-7}$ & 
    10.0$\times10^{-4}$ &
    65.0$\times10^{-4}$ &
    75.0$\times10^{-4}$ &
    101.0$\times10^{-4}$ &
    105.0$\times10^{-4}$ &
    79.0$\times10^{-4}$ &
    26.0$\times10^{-4}$ &
    6.7$\times10^{-4}$ &
    1.4$\times10^{-4}$ &
    2.8$\times10^{-5}$ \\
\hline
500 & 3.3$\times10^{-7}$ & 
    8.1$\times10^{-7}$ & 
    32.0$\times10^{-4}$ &
    41.0$\times10^{-4}$ &
    66.0$\times10^{-4}$ &
    69.5$\times10^{-4}$ &
    68.8$\times10^{-4}$ &
    19.0$\times10^{-4}$ &
    6.6$\times10^{-4}$ &
    1.4$\times10^{-4}$ &
    2.8$\times10^{-5}$ \\
\hline
600 &  2.1$\times10^{-7}$ & 
       3.5$\times10^{-7}$ & 
       11.0$\times10^{-4}$ &
       17.0$\times10^{-4}$ &
       39.0$\times10^{-4}$ &
       42.0$\times10^{-4}$ &
       58.0$\times10^{-4}$ &
       23.0$\times10^{-4}$ &
       6.3$\times10^{-4}$ &
       1.4$\times10^{-4}$ &
       2.8$\times10^{-5}$ \\
\hline
700 &  1.4$\times10^{-7}$ & 
       1.9$\times10^{-7}$ & 
       1.2$\times10^{-4}$ &
       4.1$\times10^{-4}$ &
       19.0$\times10^{-4}$ &
       22.0$\times10^{-4}$ &
       48.0$\times10^{-4}$ &
       20.0$\times10^{-4}$ &
       5.9$\times10^{-4}$ &
       1.4$\times10^{-4}$ &
       2.8$\times10^{-5}$ \\
\hline
800 &  9.0$\times10^{-8}$ & 
       1.2$\times10^{-7}$ & 
       1.3$\times10^{-7}$ & 
       3.2$\times10^{-7}$ &
       7.1$\times10^{-4}$ &
       9.1$\times10^{-4}$ &
       39.0$\times10^{-4}$ &
       18.0$\times10^{-4}$ &
       5.4$\times10^{-4}$ &
       1.3$\times10^{-4}$ &
       2.7$\times10^{-5}$ \\
\hline
900 &  6.0$\times10^{-8}$ & 
       7.8$\times10^{-8}$ & 
       1.2$\times10^{-7}$ & 
       1.4$\times10^{-7}$ & 
       1.1$\times10^{-4}$ &
       2.1$\times10^{-4}$ &
       30.0$\times10^{-4}$ &
       16.0$\times10^{-4}$ &
       5.0$\times10^{-4}$ &
       1.2$\times10^{-4}$ &
       2.6$\times10^{-5}$ \\
\hline 
\end {tabular}
\caption{The simplified model (based on Einstein Cartan gravity) cross-section measurements times branching ratios (in pb) calculated for different sets of the masses $M_{A^{\prime}}$ (in GeV), and $M_{TS}$ (in GeV), for the mass assumption given in table \ref{table:par}, with dark matter mass ($M_{\chi} = 500$ GeV), the following couplings constants $\texttt{g}_{\eta} = 0.2,~\texttt{g}_{D} = 1.2$ and at $\sqrt{s} = 13.6$ TeV.}
\label{table:tabchi}
\end{sidewaystable}

In general, the neutral dark gauge boson (A$^{\prime}$) subsequently decays to the SM fermion pairs \cite{R1}. In this article, the Einstein-Cartan gravity provides a portal for a significant production of $A^{\prime}$ dark gauge boson, which corresponds to a $U(1)_D$ symmetry in the dark sector and couples to the SM particles through only the kinetic mixing ($\epsilon \leq 10^{-2}$). For the process $pp \rightarrow TS \rightarrow \chi\bar{\chi}A^{\prime}$, if the decay channels of the $A^{\prime}$ to dark-sector particles are inaccessible, the cross section of the this process depends on the $A^{\prime}$ and torsion masses, the fermion-torsion coupling, the dark gauge boson to dark matter coupling, and the branching ratio Br($A' \rightarrow l^{+}l^{-}$), without depending on the kinetic mixing parameter.

The current analysis chooses the muonic decay of A$^{\prime}$. According to \cite{R1}, the highest significant branching ratio of A$^{\prime} \rightarrow \mu^{+}\mu^{-}$ could be reached if the following mass assumption listed in table \ref{table:par} is satisfied.

If the torsion mass ($M_{TS}$) is situated at the Planck scale, its interactions would be significantly suppressed, making it nearly undetectable in both current and future collider experiments. For torsion to have a meaningful impact on collider physics, its effective mass must be positioned at energy levels considerably lower than the Planck scale. Therefore, Einstein-Cartan gravity does not inherently necessitate a torsion field with a mass at the Planck scale. Although it is conceptually connected to the Planck scale, the effective scale of torsion is shaped by other elements within the theory, including its coupling to matter \cite{torsiomMass, R1}.
In this model, there are several free parameters, including $M_{TS}$, the mass of the dark gauge boson ($M_{A^{\prime}}$), the mass of the dark matter ($M_{\chi}$), and the coupling constants ($\texttt{g}_{\eta}$ and $\texttt{g}_{D}$). These model parameters are listed in Table \ref{table:par}. 

In the context of the dynamic torsion field, the classical value of \(\texttt{g}_{\eta} = 1/8 = 0.125\) becomes unstable due to quantum corrections, resulting in fluctuations of \(\texttt{g}_{\eta}\) as the energy scale varies \cite{R1}. Consequently, we decided to explore several values of \(\texttt{g}_{\eta}\), in addition to the classical value of 0.125. Specifically, we included one higher value (0.2) and another lower (0.061 or 0.067).

The chosen benchmark values for the coupling parameter \(\texttt{g}_{D}\) are set at 1, in line with the recommendations from the LHC Dark Matter Working Group \cite{DMrecommendations}. Additionally, a value of \(\texttt{g}_{D}\) = 1.2 is adopted based on the suggestions put forth by \cite{R1}.

Since the torsion mass falls within the TeV-scale range, the \(A^{\prime}\) is produced with a significant boost, resulting in substantial missing energy (MET) from dark-sector fermions, while the standard model (SM) backgrounds remain low \cite{R1}. 

We are currently investigating the potential production of a heavy neutral dark gauge boson ($A^{\prime}$) at the LHC, with a mass ($M_{A^{\prime}}$) that exceeds 100 GeV. The dark matter mass must be substantial according to the mass requirement specified in table \ref{table:par}. Previous searches conducted in \cite{indirect2, Nature-indirect} have indicated that there is a significant chance of detecting dark matter with masses between approximately 100 and 10000 GeV using various search techniques. As a result, we have designated the dark matter mass as $M_{\chi} = 500$ GeV.

The typical signature of this process consists of a pair of opposite-sign muons from the decay of A$^{\prime}$ plus a large missing transverse energy due to the stable dark matter $\chi$. 
The events we have studied are in the following topology: $\mu^{+}\mu^{-} + E^{miss}_{T}$. This is a clean channel for Standard Model (SM) backgrounds due to the optimization of the CMS detector for this decay channel. Additionally, according to reference \cite{zprime}, the fake rate of the QCD background was much smaller in the dimuon channels compared to the dielectron channel.
\section{Simulation of signal samples and SM backgrounds}
\label{section:MCandDat}

The SM background processes yielding muon pairs in the signal region are 
Drell-Yan (DY $\rightarrow \mu^+\mu^-$) production, 
the production of top quark pairs ($\text{t}\bar{\text{t}} \rightarrow \mu^+\mu^- + 2b + 2\nu$), $tW \rightarrow \mu^+\mu^- + 2\nu +b$, 
and production of diboson 
($W^{+}W^{-} \rightarrow \mu^+\mu^- + 2\nu$,  
$ZZ \rightarrow \mu^+\mu^- + 2\nu$ and 
$W^{\pm}Z \rightarrow \mu^\pm \mu^+\mu^- + \nu$).
The second type of background is the jet background, which comes from the misidentification of jets as muons, where a jet or multijet passes the muon selection criteria. 
This kind of background originates from two processes: W+jets and QCD multijet. 
The contamination of single and multijet backgrounds in data is usually estimated from data using a so-called data-driven method, which is explained in \cite{zprime}. Nevertheless, they are irrelevant to our study because our analysis is based on MC simulations only, and no events from W+jets pass the analysis pre-selection presented in the following section.

The signal samples, for the simplified model (based on Einstein-Cartan gravity), and the corresponding SM processes have been generated using \text{MadGraph5\_aMC@NLO} \cite{MG5} interfaced to Pythia 8 for parton shower model and hadronization \cite{R34},
and DELPHES \cite{delphes} for a fast detector simulation of the CMS experiment. 
They were generated from proton-proton collisions at the Large Hadron Collider at 13.6 TeV center of mass energy, which corresponds to the circumstances of run 3, with muon $p_T > 10$ GeV and $|\eta| < 3$ rad.

In the simplified model, we used the mass assumptions outlined in Table \ref{table:par}. In table \ref{table:tabchi}, we present the calculated measurements of cross-sections multiplied by branching ratios for the chosen benchmark values of the coupling parameters, specifically $\texttt{g}_{\eta} = 0.2$ and $\texttt{g}_{D} = 1.2$, in line with the recommendations from \cite{R1}. These calculations for the cross-sections are conducted across various sets of masses for the dark gauge boson (A$^{\prime}$) and the torsion field ($ST$).
The simulated signals, used in this analysis, are private production samples, at which we used the matrix element event generator MadGraph5 aMC@NLO v2.6.7 \cite{MG5}. 

All the Monte Carlo samples utilized in this analysis and their respective cross sections were computed in next-to-leading order and privately generated.
Thus, the contributions of the signal samples and the SM background processes have been estimated from the Monte Carlo simulations, at which they are normalized to their corresponding cross-section and integrated luminosity of 52 fb$^{-1}$ (the amount of certified data which is collected by the CMS experiment so far during LHC run 3) \cite{LumiPublicResults}.

\section{Event selection}
\label{section:AnSelection}
The selection of the event, for the analysis, has been designed to reconstruct a final state with two high transverse momentum $(p_{T})$ muons in association with missing transverse energy, accounting for the DM candidate. 
The selection is made using cuts applied to different kinematic parameters.
Each of the two muons should pass the following pre-selection: \\
$\bullet$ $p^{\mu}_{T}$ (GeV) $> 30$, \\ 
$\bullet$ $\eta^{\mu}$ (rad) $<$ 2.4, \\
$\bullet$ $\text{IsolationVarRhoCorr} < 0.1$, \\
$\text{IsolationVarRhoCorr}$ represents the isolation cut in DELPHES software
to reject muons produced inside jets.
In this cut, it is required that the scalar $p_{T}$ sum of all muon tracks within a cone of $\Delta R = 0.5$ around the muon candidate, excluding the muon candidate itself, should not exceed 10\% of the $p_{T}$ of the muon.
This cut has been corrected for the pileup effect.

Thus, each event has been selected with two opposite-charge muons, and the invariant mass of the dimuon is bigger than 60 GeV since we are looking for a resonance in the high mass regime. 

\begin{figure} [h!]
\centering
\resizebox*{9cm}{!}{\includegraphics{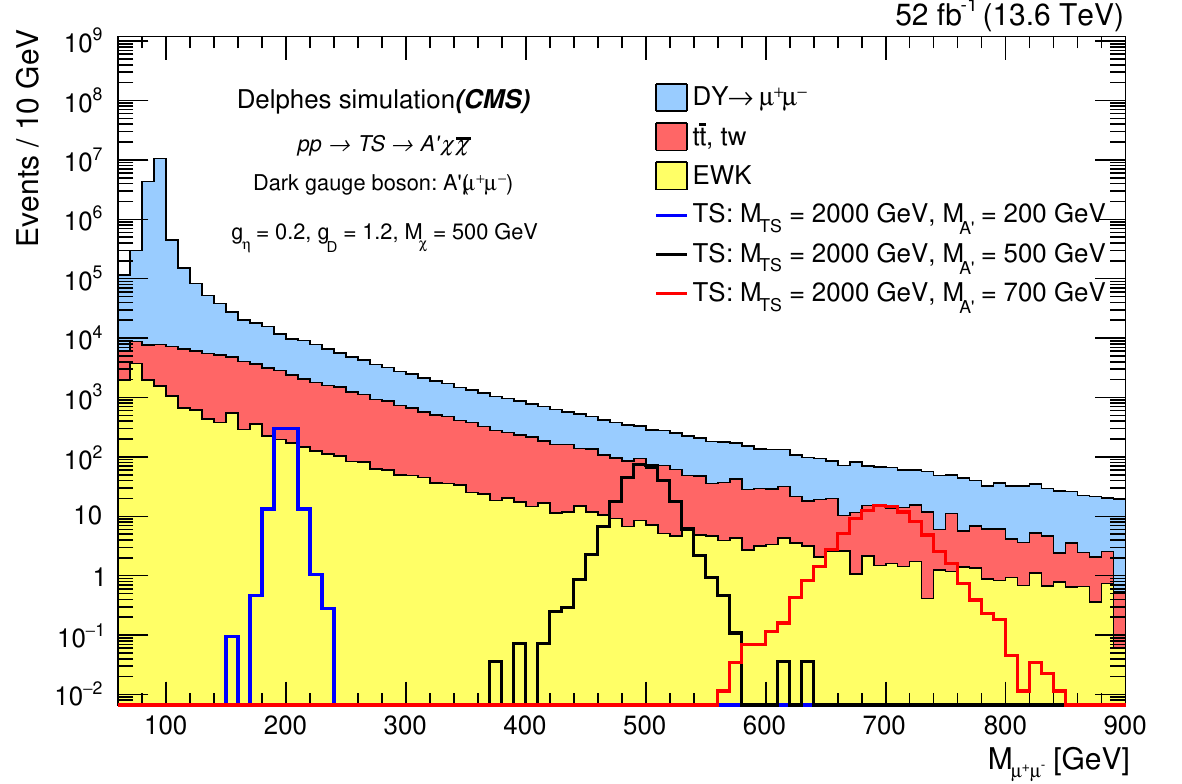}}
\caption{The measured dimuon invariant mass spectrum, for events passing selection 1 summarized in table \ref{cuts}, for the estimated SM backgrounds and different choices of dark gauge boson (A$^{\prime}$) masses generated based on the simplified model, with mass of torsion field ($M_{TS} = 2000$ GeV) and dark matter mass ($M_{\chi} = 500$ GeV).}
\label{figure:fig3}
\end{figure}
\begin{figure} [h!]
\centering
\resizebox*{9cm}{!}{\includegraphics{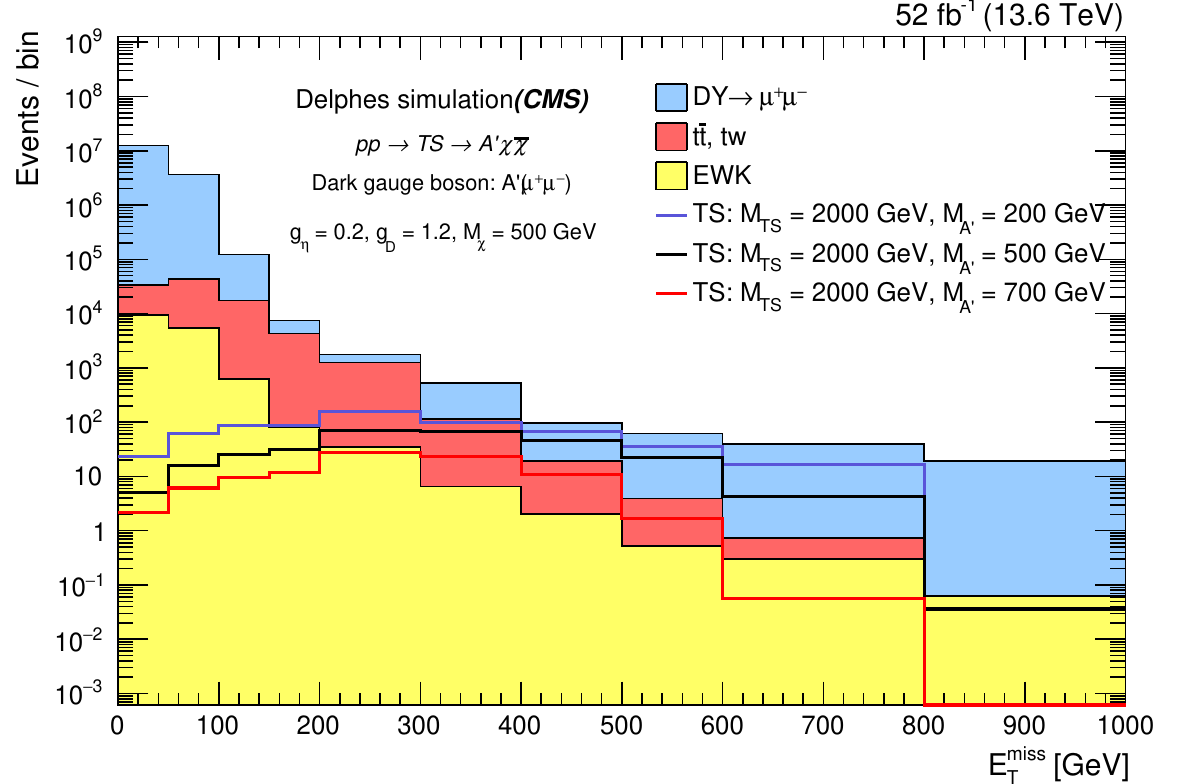}} 
  \caption{The distribution of the missing transverse energy, after selection 1; for the expected SM backgrounds, and different A$^{\prime}$ masses produced by the simplified model, with the mass of torsion field ($M_{TS} = 2000$ GeV) and dark matter mass ($M_{\chi} = 500$ GeV).}
\label{figure:fig4}
\end{figure}

Figure \ref{figure:fig3} shows the distribution of the dimuon invariant mass for events passing selection 1, summarized in table \ref{cuts}; 
The cyan histogram represents the Drell-Yan background, the yellow histogram represents the vector boson pair backgrounds (WW, WZ, and ZZ), and the red histogram represents the $t\bar{t}$ background. 
These histograms are stacked. 
While the signals of the simplified model in the framework of Einstein-Cartan gravity, which have been generated with different masses of the neutral dark gauge boson A$^{\prime}$ with fixed values of the torsion field mass ($M_{TS} = 2000$ GeV) and dark matter mass ($M_{\chi} = 500$ GeV), are represented by different colored lines, and are overlaid. 
The corresponding distribution of the missing transverse energy is presented in Figure \ref{figure:fig4}. 
It is clearly shown from these figures that the signal samples are overwhelmed by the backgrounds. So, it is necessary to apply a tighter set of cuts to discriminate signals from SM backgrounds, as will be explained in the next paragraph.

In addition to selection 1, extra-tight cuts have been designed to reduce the SM background further. 
These tight cuts are based on three variables: the first variable is related to the invariant mass of the dimuon, at which we restricted the invariant mass of the dimuon to a small range around the mass of the dark gauge boson A$^{\prime}$ to suppress backgrounds that do not peak at the A$^{\prime}$ mass, a requirement that $0.9 \times M_{A^{\prime}} < M_{\mu^{+}\mu^{-}} < M_{A^{\prime}} + 25$ as suggested in \cite{R1}. 
The second is the relative difference between the transverse momentum of dimuon $(p_{T}^{\mu^{+}\mu^{-}})$ and the missing transverse energy $(E^{\text{miss}}_{T})$, it has been selected to be less than 0.4. 
(i.e. $|p_{T}^{\mu^{+}\mu^{-}} - E^{\text{miss}}_{T}|/p_{T}^{\mu^{+}\mu^{-}} < 0.4$).
The third one is $\Delta\phi_{\mu^{+}\mu^{-},\vec{E}^{\text{miss}}_{T}}$, which is defined as difference in the azimuth angle between the dimuon direction and the missing transverse momentum direction (i.e. $\Delta\phi_{\mu^{+}\mu^{-},\vec{E}^{\text{miss}}_{T}} = |\phi^{\mu^{+}\mu^{-}}-~\phi^{miss}|$ ), it has been selected to be greater than 2.6 rad. 
These variables effectively reject DY and top quark processes. These adjustments are 
fine-tuned to maximize the sensitivity of the signal for a range of DM processes.

\begin{figure}
\centering
\subfloat[$|p_{T}^{\mu^{+}\mu^{-}}- E_{T}^{miss}| / p_{T}^{\mu^{+}\mu^{-}}$]{
  \includegraphics[width=92mm]{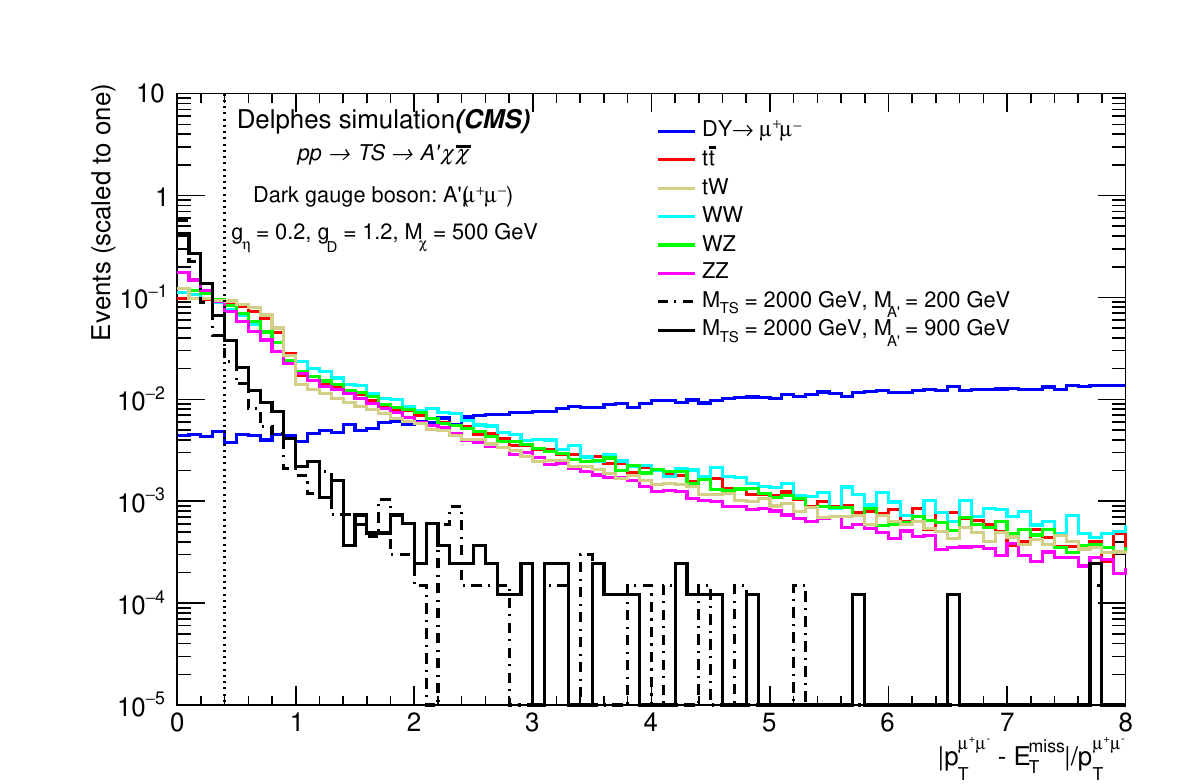}
}
\hspace{0mm}
\subfloat[$\Delta\phi_{\mu^{+}\mu^{-},\vec{E}_{T}^{\text{miss}}}$]{
  \includegraphics[width=92mm]{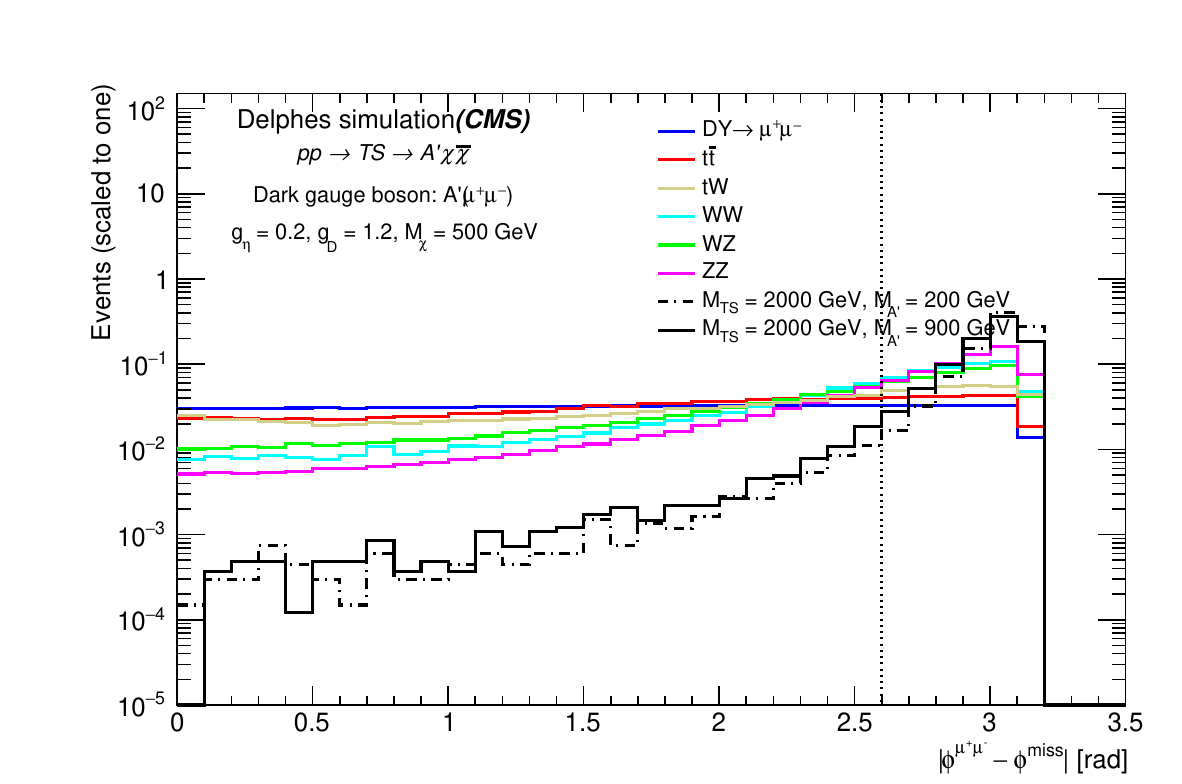}
}
\caption{Distributions of $|p_{T}^{\mu^{+}\mu^{-}} - E_{T}^{miss}| / p_{T}^{\mu^{+}\mu^{-}}$ (a) and $\Delta\phi_{\mu^{+}\mu^{-},\vec{E}_{T}^{\text{miss}}}$ (b) for two signals presentation of the simplified model corresponding to the Einstein-Cartan gravity with $M_{A^{\prime}}=$ 200 and 900 GeV and SM backgrounds, for dimuon events with each muon is passing the pre-selection cuts listed in table \ref{cuts}. The vertical dashed lines correspond to the chosen cut value for each variable. All histograms are normalized to unity to highlight qualitative features.}
\label{figure:fig70}
\end{figure}

For dimuon events passing selection 1, we present the distributions of two variables for the signal presentation of the simplified model corresponding to Einstein-Cartan gravity, along with SM backgrounds. The first variable is $|p_{T}^{\mu^{+}\mu^{-}} - E_{T}^{\text{miss}}| / p_{T}^{\mu^{+}\mu^{-}}$, shown in Figure \ref{figure:fig70}(a), while the second variable is $\Delta\phi_{\mu^{+}\mu^{-},\vec{E}_{T}^{\text{miss}}}$, shown in Figure \ref{figure:fig70}(b). The model signals were generated with dark gauge boson mass $M_{A^{\prime}}$ = 200 and 900 GeV, torsion field mass $M_{TS} = 2000$ GeV, and dark matter mass $M_{\chi} = 500$ GeV. The distributions are scaled to one. The vertical dashed lines in both figures correspond to the chosen cut value for each variable, and these tight cuts have been applied to reduce the SM backgrounds significantly.

\begin{table*}
    \centering
    \begin{tabular}{|c|c|c|}
\hline
Selection 1 & Selection 2 & Selection 3 \\
\hline
    \hline
  $p^{\mu}_{T} >$ 30 GeV &  $p^{\mu}_{T} >$ 30 GeV & $p^{\mu}_{T} >$ 30 GeV  \\
  $\eta^{\mu} <$ 2.4 rad   & $\eta^{\mu} <$ 2.4 rad & $\eta^{\mu} <$ 2.4 rad\\
     IsolationVarRhoCorr $<$ 0.1& IsolationVarRhoCorr $<$ 0.1&  IsolationVarRhoCorr $<$ 0.1\\
  $M_{\mu^{+}\mu^{-}} >$ 60 GeV  & $M_{\mu^{+}\mu^{-}} >$ 60 GeV &  $M_{\mu^{+}\mu^{-}} >$ 60 GeV\\
  &$|p_{T}^{\mu^{+}\mu^{-}}- E_{T}^{\text{miss}}|/p_{T}^{\mu^{+}\mu^{-}} <$ 0.4 &$|p_{T}^{\mu^{+}\mu^{-}}- E_{T}^{\text{miss}}|/p_{T}^{\mu^{+}\mu^{-}} <$ 0.4   \\
  &$\Delta\phi_{\mu^{+}\mu^{-},\vec{E}_{T}^{\text{miss}}} >$ 2.6 rad & $\Delta\phi_{\mu^{+}\mu^{-},\vec{E}_{T}^{\text{miss}}} >$ 2.6 rad \\
     && $0.9 \times M_{A^{\prime}} < M_{\mu^{+}\mu^{-}} < M_{A^{\prime}} + 25$\\
    \hline
        Pre-selection & Semi-final selection & Final selection \\
    \hline
    \end{tabular}
    \caption{Summary of cut-based event selections used in the analysis.}
    \label{cuts}
\end{table*}
\begin{figure} [h!]
\centering
\resizebox*{9cm}{!}{\includegraphics{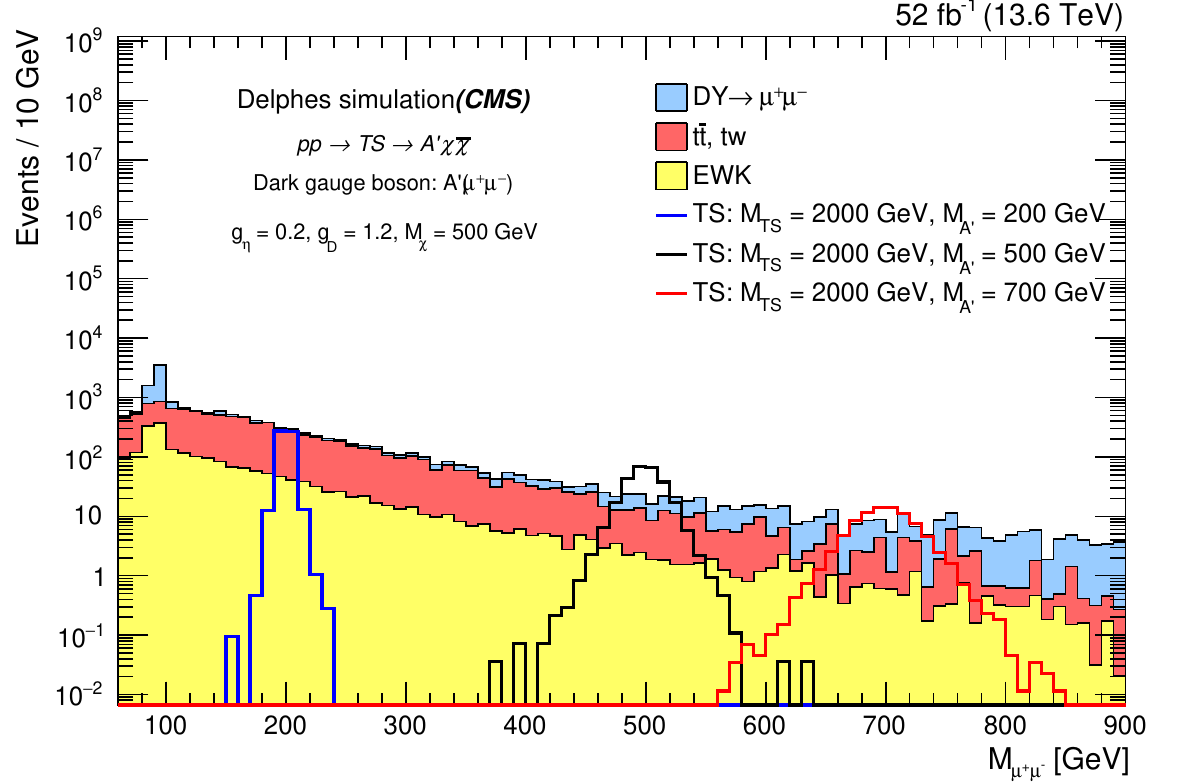}}
\caption{The dimuon invariant mass spectrum, for events passing selection 2 listed in table \ref{cuts}, for the estimated SM backgrounds and different choices of dark gauge boson (A$^{\prime}$) masses generated based on the simplified model, with mass of torsion field ($M_{TS} = 2000$ GeV) and dark matter mass ($M_{\chi} = 500$ GeV).}
\label{figure:masssemifinal}
\end{figure}

The dimuon invariant mass spectrum is shown in Figure \ref{figure:masssemifinal} for events that meet the criteria listed in Table \ref{cuts} (selection 2). 
The histograms represent the estimated SM backgrounds and various dark gauge boson (A$^{\prime}$) masses generated based on the simplified model, with a torsion field mass of $M_{TS} = 2000$ GeV and dark matter mass of $M_{\chi} = 500$ GeV.
\section{Results}
\label{section:Results}
\begin{figure}[h!]
\centering
  \resizebox*{9cm}{!}{\includegraphics{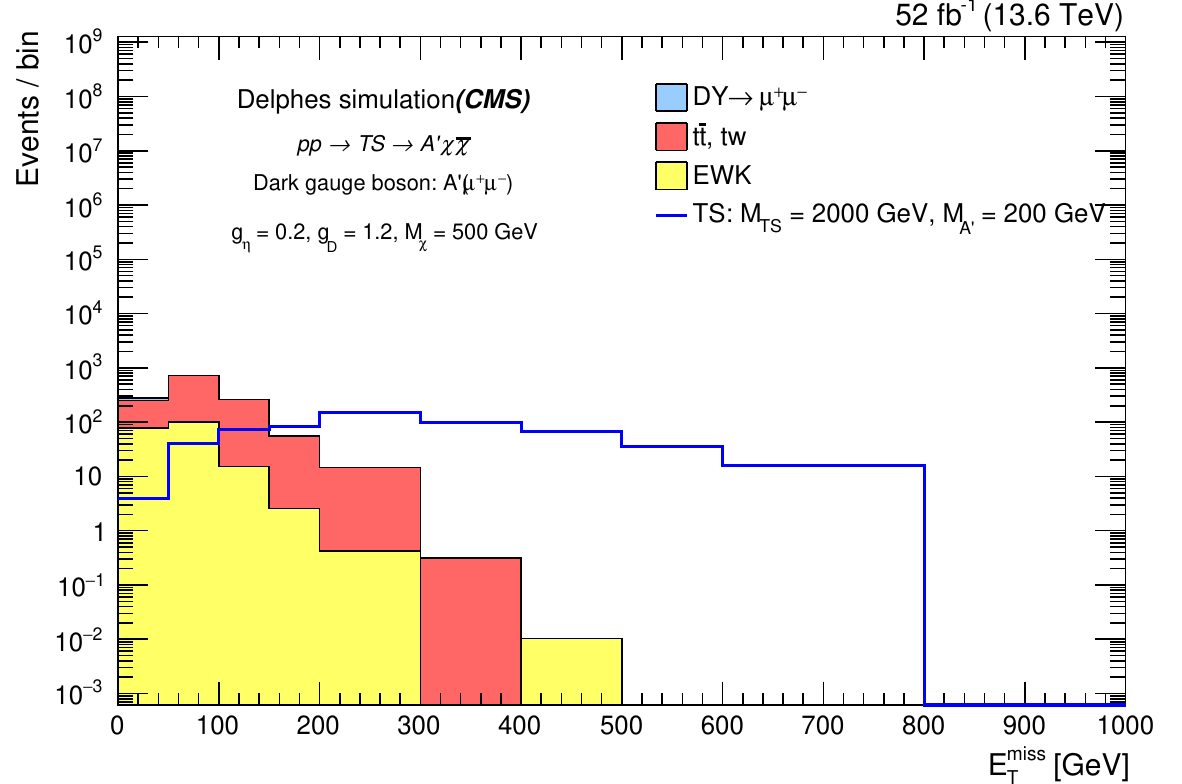}}
  \caption{The distribution of the missing transverse energy, for events passing selection 3 presented in table \ref{cuts}, for the expected SM backgrounds and one signal benchmark corresponding to the Einstein-Cartan gravity with $M_{A^{\prime}} = 200$ GeV is superimposed.}
  \label{figure:fig6}
\end{figure}

The shape-based analysis has been used based on the missing transverse energy distributions ($E^{\text{miss}}_{T}$), which are good discriminant variables since 
the signal distributions are characterized by relatively large $E^{\text{miss}}_{T}$ 
values compared to SM backgrounds.
After applying selection 3 listed in table \ref{cuts}, the distribution of the missing transverse energy is illustrated in figure \ref{figure:fig6} for the expected SM backgrounds and one signal benchmark corresponding to the Einstein-Cartan gravity with $M_{A^{\prime}} = 200$ GeV is superimposed.

To calculate the significance of the signal, we used the Asimov formula given in \cite{R2}, 
\begin{equation}
    S = \sqrt{2 \times \Big( (N_s + N_b) log\Big(1 + \frac{N_s}{N_b}\Big) - N_s\Big)},
    \label{sig:equ}
\end{equation}
$N_s$ and $N_b$ represent the number of signals and the total number of SM background events that pass the selections (3) mentioned in tables \ref{cuts}. 

\begin{figure}
\centering
  \resizebox*{8cm}{!}{\includegraphics{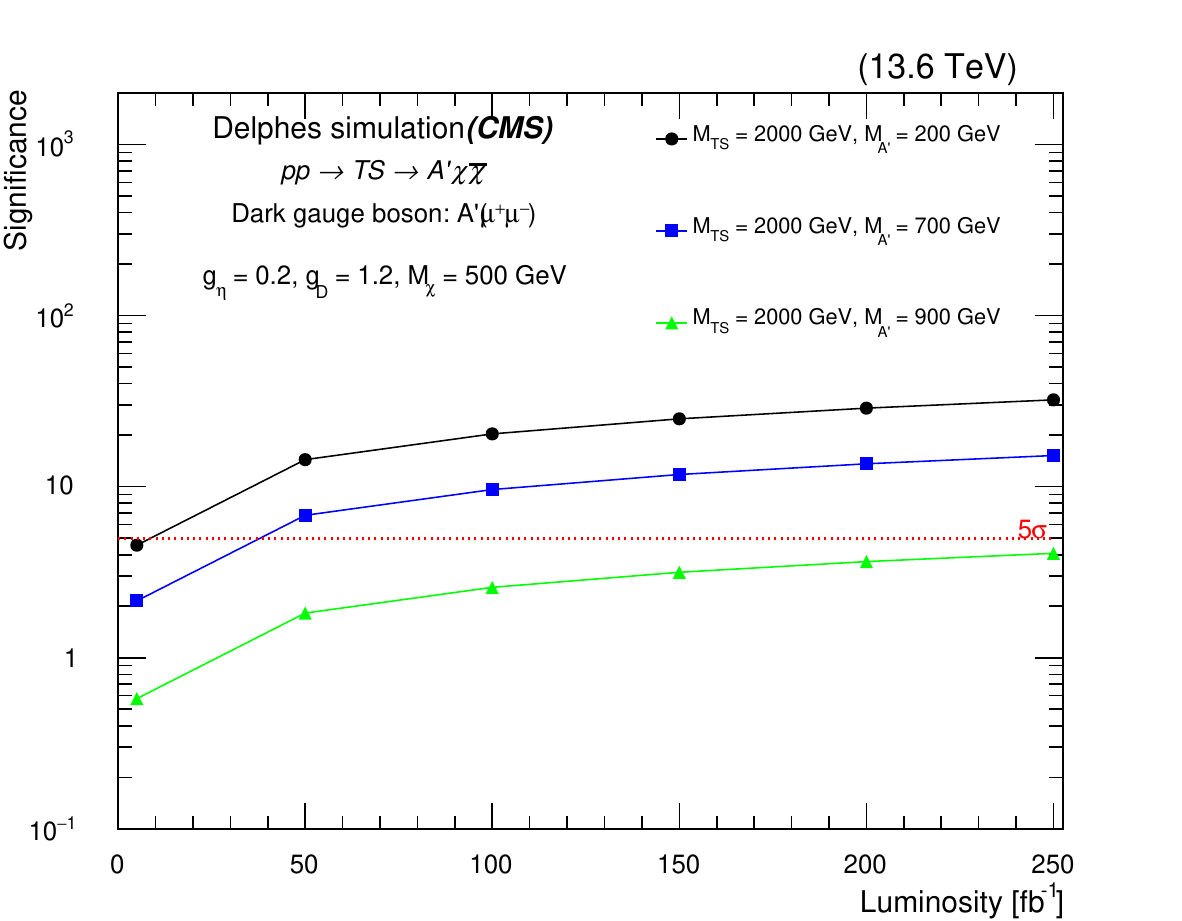}}
  \caption{The significance was presented with the integrated luminosity for Einstein-Cartan gravity, considering different dark boson masses $(M_{A'})$. The values of the coupling constants were fixed at $\texttt{g}_{\eta} = 0.2$ and $\texttt{g}_{D} = 1.2$, with $M_{TS} = 2000$ GeV, and the dark matter mass set at $M_{\chi} = 500$ GeV.
  The red dotted line refers to the $5\sigma$ discovery potential.}
  \label{figure:significance}
\end{figure}

In figure \ref{figure:significance}, the significance of A$^{\prime}$ discovery potential is plotted versus the integrated luminosity for the simplified model based on Einstein-Cartan gravity, with the dark matter mass fixed at $M_{\chi} = 500$ GeV, $M_{TS} = 2000$ GeV and the coupling constants $\texttt{g}_{\eta} = 0.2$ and $\texttt{g}_{D} = 1.2$. 
In the run3-LHC experiment, there is a possibility of detecting evidence for the torsion field with an integrated luminosity of more than 9 fb$^{-1}$ for $M_{TS} = 2000$ GeV and $M_{A^{\prime}} = 200$. Furthermore, for $M_{TS} = 2000$ GeV and $M_{A^{\prime}} = 700$, a $5\sigma$ discovery can be achieved with an integrated luminosity exceeding 38 fb$^{-1}$, as illustrated in figure \ref{figure:significance}.
However, according to the discovery potential based on the integrated luminosity, achieving a $5\sigma$ evidence is not feasible when the $M_{A'}$ masses exceed 700 GeV due to the significant decrease in the signal cross sections.
\begin{figure}
\centering
\subfloat[]{
  \includegraphics[width=80mm]{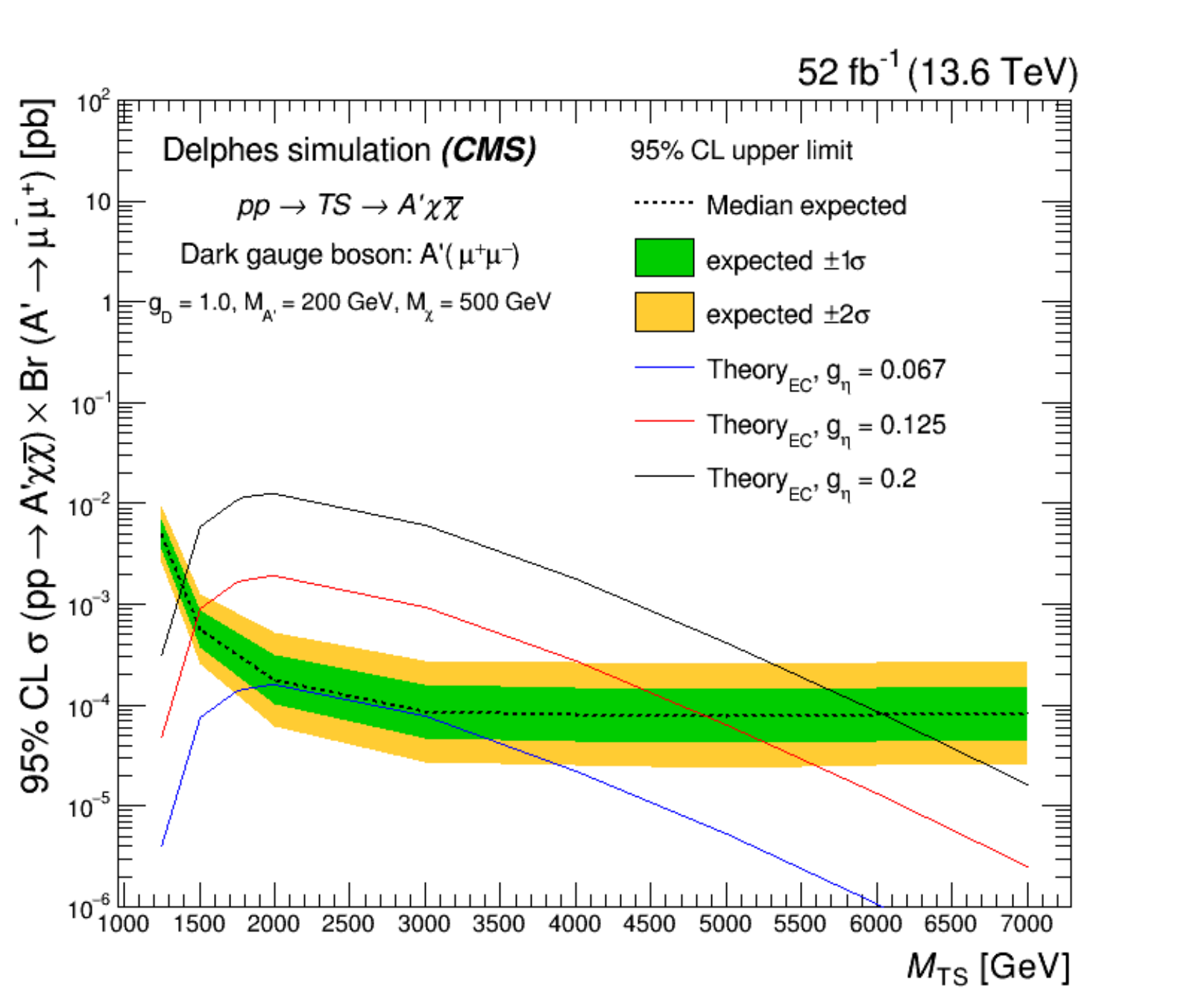}
   \label{lim1}
}
\hspace{0mm}
\subfloat[]{
  \includegraphics[width=80mm]{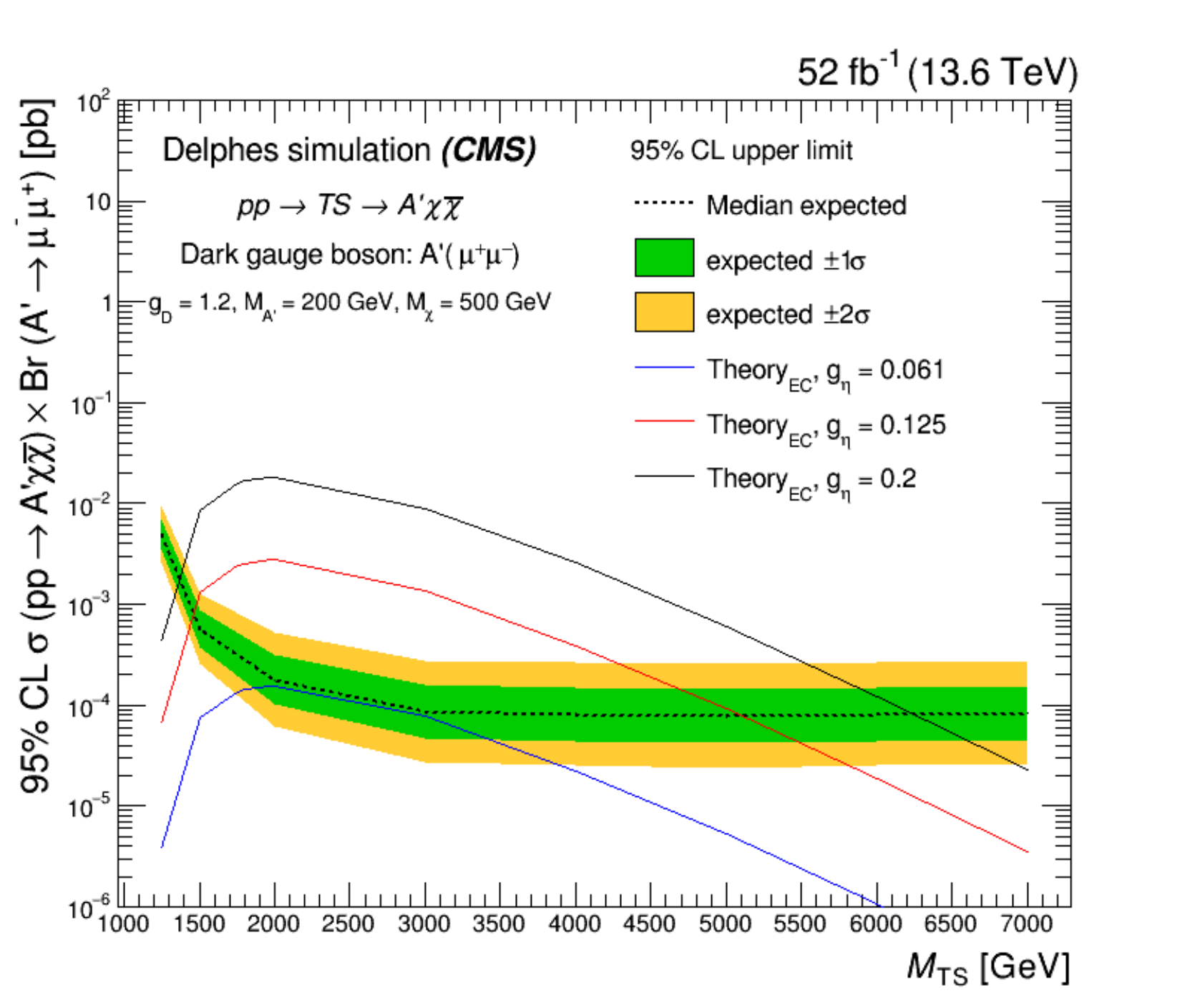}
  \label{lim2}
}
 \caption{95\% CL upper limits on the cross-section times the branching ratio (expected), as a function of the mediator's mass ($M_{\texttt{TS}}$) based on the Einstein-Cartan model for $\texttt{g}_{D} = 1.0$ \ref{lim1} and 1.2 \ref{lim2}, with the muonic decay of the A$^{\prime}$. 
  The solid colored curves illustrate the cross-section measurements multiplied by the branching ratios (in pb) for the Einstein-Cartan gravity simplified model with $\texttt{g}_{\eta} =$ 0.061 or 0.067 (in blue), 0.125 (in red), and 0.20 (in black).}
  \label{figure:fig7}
\end{figure}

To make a statistical interpretation for our results, we performed a statistical test based on the profile likelihood method, with the use of the modified frequentist construction CLs \cite{R58, R59} used in the asymptotic approximation \cite{R2} to derive exclusion limits on the product of signal cross sections and branching fraction Br($A^{\prime}$ $\rightarrow \mu\mu$) at 95\% confidence level. 
The limits and significances presented have an adhoc flat 10\% uncertainty, systematic uncertainties taken into account to cover all possible systematic effects.

The 95\% upper limit on the product of the cross-section and the branching ratio, plotted against the mass of the torsion field \(M_{TS}\), is shown in figure \ref{figure:fig7}. This analysis is based on a simplified model using Einstein-Cartan gravity and incorporates the muonic decay of the \(A^{\prime}\) for coupling constants \(\texttt{g}_{D} = 1.0\) \ref{lim1} and \(1.2\) \ref{lim2}. In this context, we consider a dark matter mass of \(M_{\chi} = 500\) GeV and \(M_{A^{\prime}} = 200\) GeV.
The solid curves illustrate the cross-section measurements multiplied by the branching ratios (in pb) for a simplified model with $\texttt{g}_{\eta} =$ 0.061 or 0.067 (in blue), 0.125 (in red) and 0.20 (in black). These measurements correspond to various values of the torsion field mass ($M_{TS}$), as detailed in Table \ref{table:tabchi}, and are calculated at a center-of-mass energy of $\sqrt{s} = 13.6$ TeV.

In Figure \ref{figure:fig7}, the intersections of the solid-colored curves with the median expected curve, represented as a dotted line, indicate the mass ranges where the production of the torsion scalar (TS) can be excluded. For a coupling constant of \(\texttt{g}_{\eta} = 0.125\), this exclusion range is between 1476 and 4860 GeV. For \(\texttt{g}_{\eta} = 0.2\), the range extends from 1387 to 6040 GeV when \(\texttt{g}_{D} = 1.0\).
Similarly, when \(\texttt{g}_{D} = 1.2\), the ranges are as follows: from 1460 to 5100 GeV for \(\texttt{g}_{\eta} = 0.125\) and from 1369 to 6241 GeV for \(\texttt{g}_{\eta} = 0.2\). 

Our findings show that \(\texttt{g}_{\eta}\) must be greater than 0.067 for \(\texttt{g}_{D}\) = 1 and 0.061 for \(\texttt{g}_{D}\) = 1.2.

These findings are summarized in Table \ref{limit-values}.
As a result, when we set \(M_{\chi} = 500\) GeV and \(M_{A^{\prime}} = 200\) GeV, we find that larger values of the coupling constant \(\texttt{g}_{\eta}\) (such as 0.2) lead to the exclusion of a wider range of torsion field mass values. On the other hand, when the \(\texttt{g}_{D}\) coupling is increased from 1.0 to 1.2, the outcome remains relatively unchanged.

\begin{table}
    \centering
    \begin{tabular}{|c|c|c|}
\hline
$\texttt{g}_{D}$ & $\texttt{g}_{\eta} = 0.125$ & $\texttt{g}_{\eta} = 0.2$ \\
    \hline
  1.0 & 1476 - 4860 GeV  & 1387 - 6040 GeV  \\
  \hline
  1.2 & 1460 - 5100 GeV  & 1369 - 6241 GeV \\
   \hline
    \end{tabular}
    \caption{The expected limit values at 95\% CL for different values of the model coupling constants $\texttt{g}_{D}$ and $\texttt{g}_{\eta}$.}
    \label{limit-values}
\end{table}

\begin{figure}
\centering
  \resizebox*{8.5cm}{!}{\includegraphics{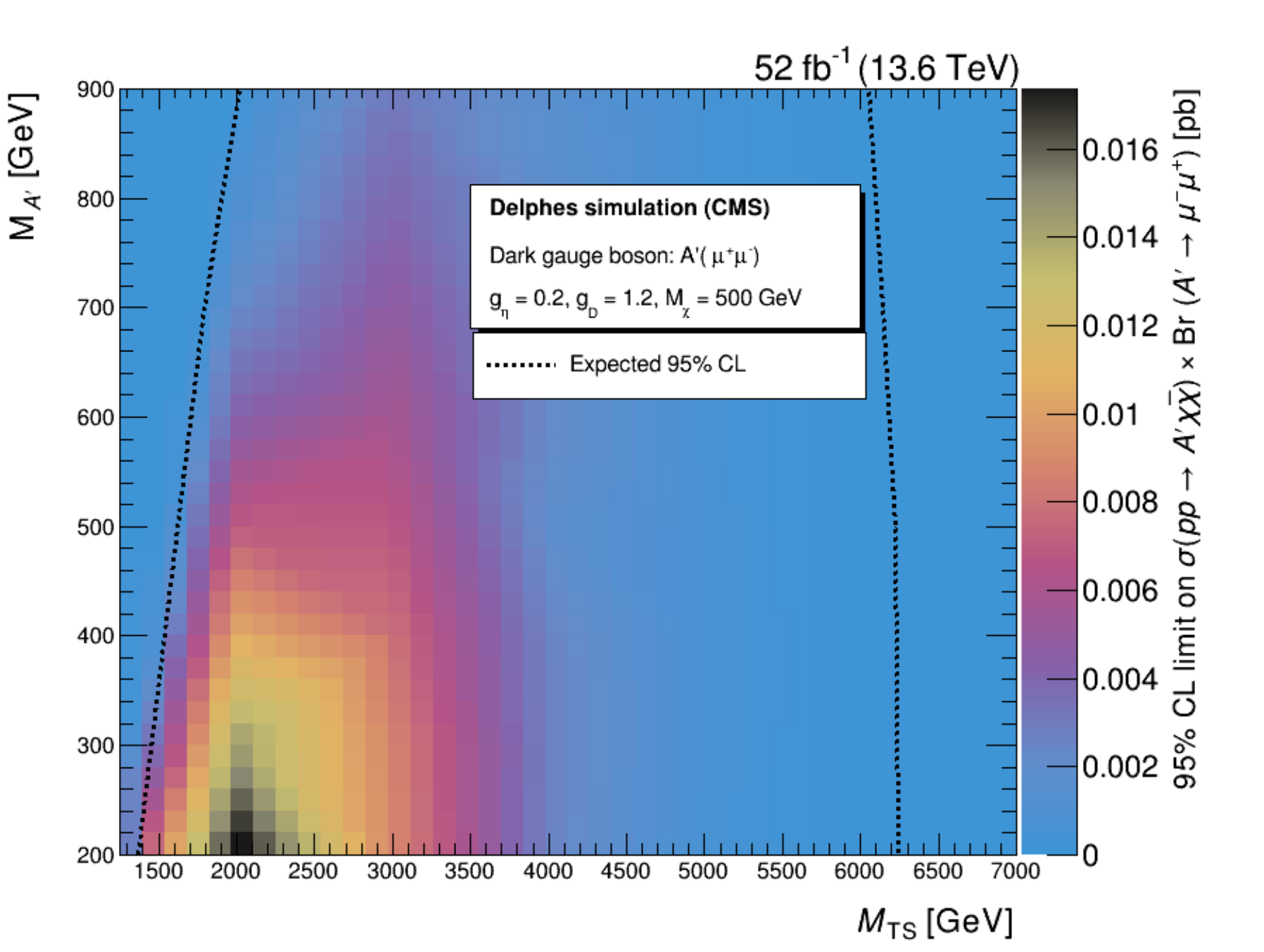}}
  \caption{The 95\% CL upper limits on the product of the cross-section and branching fraction from the inclusive search, for variations of pairs of the simplified model parameters ($M_{\texttt{TS}}$ and $M_{A^{\prime}}$). The filled region indicates the upper limit. The dotted black curve indicates the expected exclusions for the nominal A$^{\prime}$ cross-section.}
  \label{figure:fig8}
\end{figure}

In the context of the dynamic torsion field, the classical value of \(\texttt{g}_{\eta} = 1/8\) becomes unstable due to quantum corrections, leading to variations in \(\texttt{g}_{\eta}\) with changes in the energy scale \cite{R1}. Consequently, we have opted for \(\texttt{g}_{\eta} = 0.2\). Thus, in our simplified model, which is based on Einstein-Cartan gravity and features coupling constants of \(\texttt{g}_{D} = 1.2\), \(\texttt{g}_{\eta} = 0.2\), and a dark matter mass of \(M_{\chi} = 500\) GeV, we present the cross-section times the branching ratio limit in Figure \ref{figure:fig8}. This is illustrated as a function of the mediator masses \(M_{TS}\) and the masses of the dark neutral gauge boson \(M_{A^{\prime}}\).
The region between the respective pair of the expected 95\% dotted line is excluded.
The results from the inclusive signal regions exclude expected values of up to $1369 < M_{TS} < 6241$ GeV. 
\section{Summary}
\label{section:Summary}
We have proposed a methodology for searching for the torsion field (TS) at the LHC, which can decay into dark matter (DM) pair particles. One of these DM particles is sufficiently heavy to decay into dark neutral gauge bosons (A$^{\prime}$) along with another DM particle. This study has been conducted within the framework of the $U(1)_{D}$ simplified model, based on Einstein-Cartan gravity.

The presented analysis has been performed using the simulated proton-proton collisions
corresponding to the LHC run 3 with 13.6 TeV center of mass energy, for an integrated
luminosity of 52 fb$^{-1}$, which corresponds to the amount of certified data collected by the CMS experiment so far during LHC run 3. 

Results from the muonic decay mode of A$^{\prime}$ are discussed, with fixing the values of the dark matter mass ($M_{\chi} = 500$ GeV). We have considered the variations of several parameters of the signal model: the torsion field mass $(M_{TS})$ and the dark neutral gauge boson mass $(M_{A^{\prime}})$ in addition to the model coupling constants \(\texttt{g}_{\eta}\) and \(\texttt{g}_{D}\).

The final analysis indicates that achieving a $5\sigma$ discovery of the A$^{\prime}$ dark gauge boson in the $\mu^+ \mu^-$ decay channel is feasible. This can be accomplished for $M_{A^{\prime}}$ values ranging from 200 to 700 GeV, given $M_{TS} = 2000$ GeV, with an integrated luminosity of less than 40 fb$^{-1}$ at the 13.6 TeV center of mass energy of the LHC run 3.

In our current analysis, we conducted limit calculations at a 95\% confidence level for both the cross section and the torsion mass (\(M_{TS}\)), specifically with values set at \(M_{\chi} = 500\) GeV and \(M_{A^{\prime}} = 200\) GeV. We not only established limits on the cross section and torsion mass but also examined a variety of values for the model coupling constants, \(\texttt{g}_{\eta}\) and \(\texttt{g}_{D}\). 

Our results show that \(\texttt{g}_{\eta}\) needs to be greater than 0.067 for \(\texttt{g}_{D}\) = 1 and 0.061 for \(\texttt{g}_{D}\) = 1.2; 
notably, higher values of this coupling constant, such as 0.2, lead to the exclusion of a wider range of torsion field mass values. Furthermore, our findings suggest that adjusting \(\texttt{g}_{D}\) to either 1.0 or 1.2 results in only a minimal effect on the limit calculations. 

If no new physics emerges, the comprehensive analysis, which is based solely on event-level kinematic variables, excludes models within the mass range of \(1369 < M_{TS} < 6241\) GeV at a 95\% confidence level. This applies to \(M_{A^{\prime}}\) values spanning from 200 to 900 GeV, given that \(\texttt{g}_{\eta}\) is set to 0.2 and \(\texttt{g}_{D}\) is at 1.2.

\begin{acknowledgments}
The author of this paper would like to thank Cao H. Nam, the author of \cite{R1}, for his useful discussions about the theoretical models, and for sharing with us the Universal FeynRules Output (UFO) for the model that was used for the generation of the events. 
This paper is based on works supported by the Science, Technology, and Innovation Funding Authority (STDF) under grant number 48289.
\end{acknowledgments}

\textbf{Data Availability Statement:} This manuscript has no associated data or the data will not be deposited. 


\end{document}